\documentclass[pra,aps,amssymb,twocolumn,showpacs]{revtex4}
\usepackage{graphicx}
\newcommand{\half}{{\frac{1}{2}}}
\newcommand{\ket}[1]{|#1\rangle}

\newcommand{\mr}{\mathrm}

\begin{document}

\title{Enhancement of entanglement transfer in a spin chain by phase shift control}
\author{Koji Maruyama$^{1,2}$, Toshiaki Iitaka$^3$, and Franco Nori$^{1,4}$}
 \affiliation{$^1$Frontier Research System, RIKEN (The Institute of Physical and Chemical
Research), Wako-shi 351-0198, Japan\\
$^2$Laboratoire d'Information Quantique and QUIC, CP 165/59, Universit\'{e} Libre de
Bruxelles, 1050 Bruxelles, Belgium\\
$^3$Computational Astrophysics Laboratory, RIKEN (The Institute of Physical and Chemical
Research), Wako-shi 351-0198, Japan\\
$^4$ Center for Theoretical Physics, Physics Department, Center for the Study of Complex Systems,
University of Michigan, Ann Arbor, MI 48109-1040, USA}
\date{\today}

\begin{abstract}
We study the effect of a phase shift on the amount of transferrable two-spin entanglement in a
spin chain. We consider a ferromagnetic Heisenberg/XY spin chain, both numerically and
analytically, and two mechanisms to generate a phase shift, the Aharonov-Casher effect and the
Dzyaloshinskii-Moriya interaction. In both cases, the maximum attainable entanglement is shown to
be significantly enhanced, suggesting its potential usefulness in quantum information processing.
\end{abstract}

\maketitle

\section{Introduction}
Transferring quantum information reliably and efficiently is an important task in quantum
information processing. For example, quantum communication protocols, such as quantum key
distribution (or quantum cryptography), usually require two (or more) distant parties to share
entanglement of high quality to achieve tasks that are impossible in the regime of classical
mechanics. Also in a typical situation we encounter in the standard quantum computation model, we
need to couple two spatially separated qubits in order to perform two qubit unitary operations,
e.g., a controlled-NOT gate.

Most common approaches to this task include methods with an information bus, guided ions (atoms),
flying photons, a sequence of swapping operations between neighboring qubits, etc. However, these
methods require additional complexity in structures, manipulations and controls of the interaction
between qubits, as well as repeated conversions between the qubit state and another physical
degree of freedom. Flying photons may be the best information carrier over macroscopic distances,
but may not be so for microscopic scales of the order of, say, a few micrometers. This is why
there has been intensive research activity in the past few years on quantum information transfer
via arrays/chains of stationary qubits that are interacting with their neighboring qubits.

Typically, in previous studies of this topic, the quantum information channel consists of one or
more chains of spin-1/2 particles, each of which interacts with their nearest neighbors. The
interaction between spins can be described by either the Heisenberg or the XY (or variations of
these, e.g., the XXZ) model with some relevant parameters for coupling strengths, anisotropy, etc.
These types of models attract much attention because they are in principle sufficient for
implementing quantum information processing: examples of proposed methods are those with quantum
dots and particles trapped in an optical lattice.

A number of important and interesting results have been reported in this research area: for
example, sending quantum information through a spin chain without modulation \cite{sougato03},
entanglement transport with an anisotropic XY model \cite{subrahmanyam04}, perfect transfer by
manipulating the coupling strengths \cite{albanese04,christandl05}, near perfect transfer with
uniform couplings by a spatially varying magnetic field \cite{shi05}, Fourier analysis-based
quantum information encoding \cite{osborne04}, measurement-assisted transfer through two parallel
chains \cite{burgarth05}, specific realizations of the method in \cite{sougato03} with
superconducting qubit array \cite{romito05,lyakhov05}, perfect transfer by local measurements on
individual spins \cite{verstraete04}, and also transfer through a chain of coupled harmonic
oscillators \cite{plenio04}.

The geometry of the spin chain could be more general in principle, but we will primarily consider
a one-dimensional chain or ring of $N$ spin-1/2 particles for simplicity. If we look at the chain
as a whole, we can think of independent quantum states of collective modes, i.e., eigenstates of
the Hamiltonian for the whole chain. The propagation of quantum information encoded in a spin can
be thought of as the interference between all modes, which evolve independently.

Hence, a naive strategy towards quantum information transfer of better quality would be to control
the propagation of each mode, which will affect the (constructive) interference at a certain
target site. Here we take this approach and consider the effect of changes in the energy spectrum
and dispersion relation of the mode (or `spin wave'), that induce a spin current in the chain. As
a result of the induced change in the energy spectrum, there will also be a change in the time
evolution of each mode and thus the interference between these modes. An advantage of this
approach is that the pairwise coupling strengths $J_{ij}$ between $i$-th and $j$-th spins and the
external electromagnetic field can be taken as a constant over the whole chain, unlike some
schemes proposed before. That is, they do not have to be manipulated site by site, regardless of
the starting and target sites for the transfer.

In this paper, we study the effect of a phase shift on the amount of transferrable entanglement
through a spin chain. We shall primarily focus on the transfer of entanglement, rather than the
state itself, since entanglement is the key to achieving highly non-classical information
processing. Besides, keeping the fidelity of the (entangled) two-spin system is harder than
keeping the fidelity of a single spin. This can be stated more precisely as ``the entanglement
fidelity associated with a trace-preserving map is lower than or equal to the corresponding
input-output fidelity of a subsystem" \cite{schumacher96}. In short, attaining an entanglement
transfer of high quality guarantees a state transfer of high quality.

Further, we assume that the whole system is in the `one-magnon' state, in which the total number
of up spins in the chain is one. This situation is simple enough to start our analysis with and is
indeed reasonable when considering information transfer. Thus, all spins in the initial state,
except the one that is the subject of the transfer operation, are initially down. As for the
interaction between spins, we will consider the standard isotropic Heisenberg model. The
Heisenberg model is actually equivalent to the XY model under the one-magnon assumption, therefore
our analysis will be applicable to a wide range of physical systems. Also, as assumed (mostly
tacitly) in the literature listed above, we make another assumption not to let useful entanglement
pass by the target site: the entanglement at the target site(s) can be extracted later on at will
or can be retained for further operations, including entanglement distillation \cite{horodecki97},
to maximize the efficiency of subsequent processes.

We will consider two different ways of generating phase shifts. One is due to the Aharonov-Casher
effect \cite{aharonov84}, while the other is induced by the Dzyaloshinskii-Moriya interaction
\cite{dzyaloshinskii58,moriya60}. There are other means to generate a phase shift, or equivalently
a (persistent) spin current, in a chain/ring, such as those reported in
\cite{schutz03,schmeltzer05}. However, we focus on the two above because these seem to be
sufficient to demonstrate a significant entanglement-enhancing effect due to phase shifts. We
discuss only chains of ring geometry because the effect is absent in open ended chains as shown in
the Appendix.

\section{Pairwise entanglement in spin chains}
Let us start with a description of the entanglement between an arbitrary pair of spins in an
$N$-spin chain within the one-magnon condition. The properties of the spin chain, such as its
geometry, the nature of the interaction, etc. could be any at this point. As there is only one up
spin in total, the state of the whole chain at time $t$ can be given by
\begin{equation}\label{one_magnon_state}
\ket{\Psi(t)}=\sum_j\alpha_j(t) S_j^+ \ket{0}^{\otimes N},
\end{equation}
where $S_j^+=S_j^x+iS_j^y$ is the raising operator defined with the spin-1/2 operators $S_j^\alpha
\hspace{0.5mm} (\alpha=x,y,z)$ for the $j$-th spin. The lowering operator is defined as its
Hermite conjugate, $S_j^-=S_j^x-iS_j^y$. Throughout this paper, we will let $\ket{0}$ denote the
spin-down state, while $\ket{1}$ is the spin-up state. As in standard entanglement transfer
scenarios with the one-magnon assumption, the magnon is initially localized at a single site. With
such initial conditions, we can identify the amplitudes $\alpha_j(t)$ as the propagators, or the
Green functions, from the point of view of wave mechanics. They have all information to describe
the time evolution of a magnon wave packet.

In order to evaluate the pairwise entanglement in $\ket{\Psi(t)}$, we employ the concurrence
\cite{wootters98} as its measure. The concurrence $C$ in a bipartite state $\rho$ is defined as
\begin{equation}\label{def_concurrence}
C:=\max \{0, \lambda_1-\lambda_2-\lambda_3-\lambda_4\},
\end{equation}
where $\lambda_i$ are the square roots of the eigenvalues of matrix $R$ in descending order. The
matrix $R$ is given as a product of $\rho$ and its time-reversed state, namely
\begin{equation}
R=\rho (\sigma^y\otimes\sigma^y)\rho^* (\sigma^y\otimes\sigma^y),
\end{equation}
where $\sigma^y$ is one of the standard Pauli matrices and the star denotes the complex conjugate.
The concurrence takes its maximum value 1 for the maximal entanglement, while it is 0 for all
disentangled qubits.

We will compute the concurrence $C_{l_1,l_2}(t)$, between the $l_1$-th and $l_2$-th spins, at time
$t$ by tracing out all spins except those two. Then, $C_{l_1,l_2}(t)$ can be written as
\begin{equation}\label{conc_general}
C_{l_1,l_2}(t) = 2 |\alpha_{l_1}(t)|\cdot |\alpha_{l_2}(t)|.
\end{equation}

Hence Eq. (\ref{conc_general}) states that the concurrence between the two sites in a chain can be
expressed as a product of the absolute values of propagators. That the entanglement is determined
by the propagators supports our approach towards enhancing the entanglement by phase shift,
because the propagators are naturally affected by change in dispersion relation caused by phase
shifts.

\section{Entanglement transfer with the Aharonov-Casher effect}\label{ac-type}
First, we consider the Aharonov-Casher effect \cite{aharonov84} as a physical mechanism that
causes a phase shift in the collective modes. When a neutral particle with magnetic moment
$\vec{\mu}$ travels from $\vec{r}$ to $(\vec{r}+\Delta\vec{r})$ in the presence of electric field
$\vec{E}$, the wave function of the particle acquires an extra phase, which is the Aharonov-Casher
(AC) phase,
\begin{equation}\label{ac_phase}
\Delta\theta=\frac{1}{\hbar c^2}\int_{\vec{r}}^{\vec{r}+\Delta\vec{r}} \vec{\mu}\times
\vec{E}(\vec{x})\cdot d\vec{x},
\end{equation}
in addition to the ordinary dynamical phase. The physical origin of the AC effect is that a
particle moving in an electric field feels a magnetic field as well due to relativistic effects:
the AC effect is essentially equivalent to spin-orbit coupling. If there is no external field
applied to the ring of spins, the dispersion relation should be symmetric with respect to the zero
wave number ($k=0$), i.e. $E_k=E_{-k}$, due to the rotational symmetry of the system. However, if
the accumulated AC phase along a ring does not vanish after a $2\pi$ rotation, the dispersion
relation will change as the applied field breaks the (spatial) symmetry. Consequently, the
propagation speed of each mode will be affected and the concurrence between any two sites can be
expected to change accordingly.

\begin{figure}
\begin{center}
\includegraphics[scale=.35]{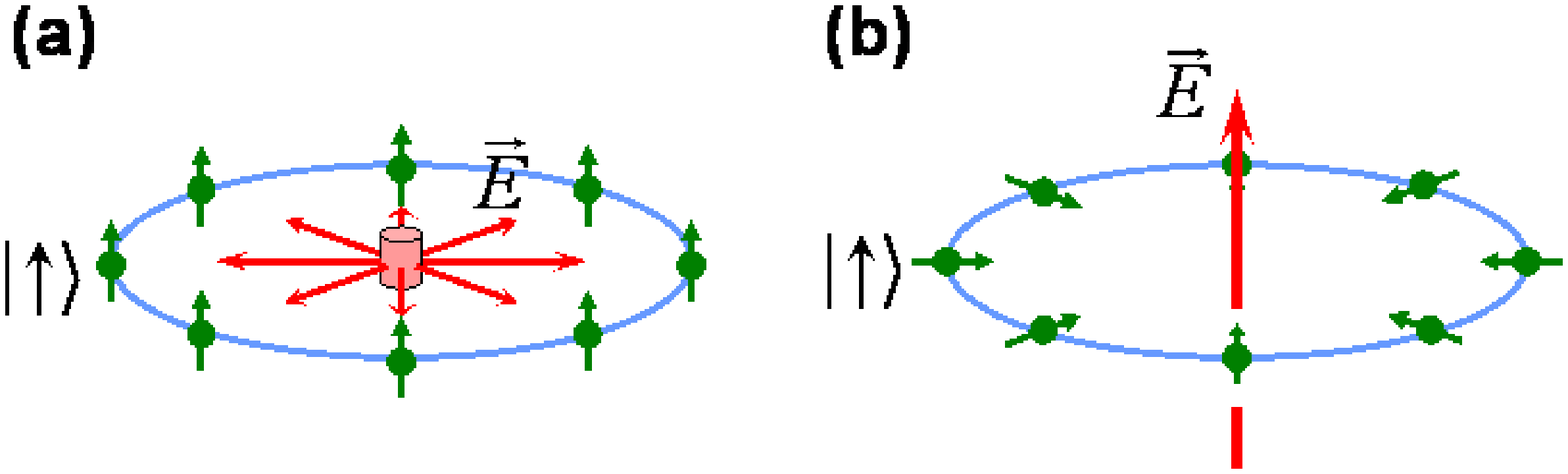}
\caption{Examples of configurations for the Aharonov-Casher effect in a ring-shaped spin chain. In
(a), the $z$-axis is taken to be parallel to the direction perpendicular to the plane that
contains the spin chain. With an electric field $\vec{E}$, which is directed to the radial
direction, the term $\vec{\mu}\times \vec{E}(\vec{x})\cdot d\vec{x}$ in Eq. (\ref{ac_phase}) takes
its largest value. Alternatively, as in (b), the directions for the spin and the electric field
can be swapped to have the same $\vec{\mu}\times \vec{E}(\vec{x})$. The magnetic moment of spin
eigenstate, $\ket{\uparrow}$ or $\ket{\downarrow}$, is parallel to the radial direction.}
\label{ac_effect}
\end{center}
\end{figure}

Figure \ref{ac_effect} sketches two possible configurations to have the AC phase effectively. The
geometry in Fig. \ref{ac_effect}(a) is very similar to that for electrons in an atom. An electric
field diverges radially, and the $z$-axis is taken to be perpendicular to the plane containing the
ring. The term $\vec{\mu}\times \vec{E}(\vec{x})\cdot d\vec{x}$ in Eq. (\ref{ac_phase}) takes its
largest value when a (quasi-) magnetic moment (magnon) travels along the chain. The electric field
could be generated by, for example, a charge on a wire at the center of the ring. Alternatively,
the directions for the spin and the electric field can be swapped as in Fig. \ref{ac_effect}(b) to
have the same $\vec{\mu}\times \vec{E}(\vec{x})$.

\begin{figure}
\begin{center}
\includegraphics[scale=.47]{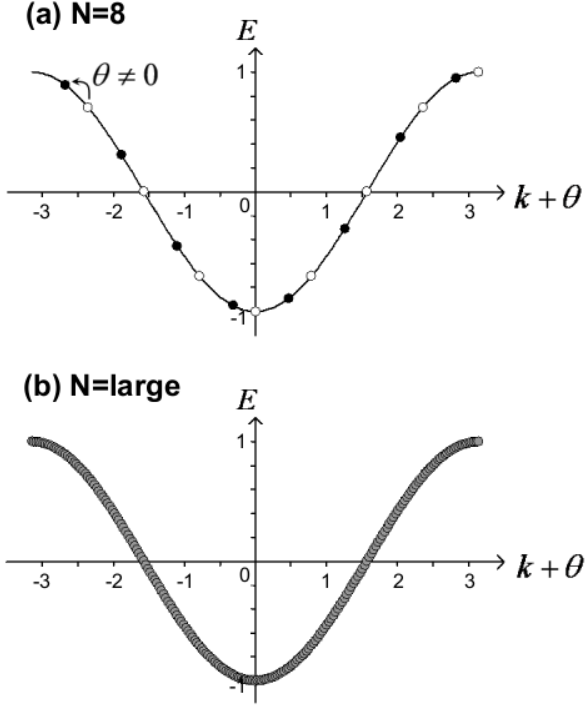}
\caption{Energy spectrum $E_k=-\cos (k+\theta)$ (in the units of $J$). (a) A non-zero $\theta$
changes the dispersion relation represented with open circles (no phase shift) to the one with
filled circles ($\theta\ne 0)$. The number of sites $N$ is taken to be 8. (b) The energy spectrum
when $N$ is large.} \label{eng_spectrum}
\end{center}
\end{figure}

We consider a Heisenberg chain of $N$ spin-1/2 particles interacting ferromagnetically with their
nearest neighbors. The phase acquisition due to the AC effect modifies the standard Heisenberg
model Hamiltonian to \cite{cao97,meier03}
\begin{eqnarray}\label{heisenberg_hamiltonian}
H &=& -\sum_{j=1}^{N} \left[\half \left(e^{i\theta} S_j^+ S_{j+1}^- + e^{-i\theta} S_j^-
S_{j+1}^+ \right) + S_j^z S_{j+1}^z \right. \nonumber \\
& & \left. + h S_j^z  \right],
\end{eqnarray}
where the interaction strength is taken as $J=-1$ for all neighboring pairs and $h$ is the
magnetic field, which is taken to be uniform and parallel to the $z$-direction. The phase change
$\theta$ between neighboring spins is given by Eq. (\ref{ac_phase}) with $\vec{r}=\vec{r}_j$ and
$\Delta\vec{r}=\vec{r_{j+1}}-\vec{r_j}$. The ring-shaped configuration is represented by periodic
boundary conditions, i.e., $N+1=1$. The Hamiltonian $H$ can be diagonalized with the help of the
Jordan-Wigner transformation \cite{lieb61,pfeuty70} that maps spins under $H$ to spinless
fermions. The annihilation and creation operators for the fermion at site $j$ are
\begin{eqnarray}\label{jw_transformation}
c_j &=& \exp\left(\pi i \sum_{l=1}^{j-1}S_l^+ S_l^- \right) S_j^- \nonumber \\
\mr{and}\hspace{2mm} c_j^\dagger &=& S_j^+ \exp\left(-\pi i \sum_{l=1}^{j-1}S_l^+ S_l^- \right).
\end{eqnarray}
Under the one-magnon condition, the Hamiltonian $H$ is now diagonalized as
\begin{eqnarray}\label{diag_h}
H &=& - \sum_{j=1}^{N} \left[\half \left(e^{i\theta}c_j^\dagger c_{j+1} +
e^{-i\theta}c_{j+1}^\dagger c_j
\right) \right. \nonumber \\
& & \left. -\half \left( c_j^\dagger c_j +c_{j+1}^\dagger c_{j+1} \right)+h\left( c_j^\dagger c_j -\half \right) +\frac{1}{4} \right] \nonumber \\
&=& \sum_k E_k \eta_k^\dagger \eta_k
\end{eqnarray}
with a further linear transformation $\eta_k=\sum_j \phi_{kj}^* c_j$. The energy eigenvalues $E_k$
can be computed as
\begin{equation}\label{energy_spectrum}
E_k= -\cos (k+\theta) + \left(1-\frac{N}{4}\right)-h \left(1-\frac{N}{2}\right),
\end{equation}
where $k=2\pi n/N$ with $-N/2< n \le N/2$, and $\phi_{kj}=1/\sqrt{N}e^{ikj}$. As the second and
third terms of Eq. (\ref{energy_spectrum}) are constant, we will omit them hereafter. A one-magnon
eigenstate can be obtained accordingly with the form of $\eta_k^\dagger$ as
\begin{equation}\label{eigen_ring}
\ket{k} := \eta_k^\dagger\ket{0}^{\otimes N}= \frac{1}{\sqrt{N}}\sum_j e^{ikj} S_j^+
\ket{0}^{\otimes N}.
\end{equation}
The presence of the extra phase $\theta$ is reflected only in the energy spectrum, while the
expression for eigenstates is unchanged. The change in the dispersion relation by a phase shift is
illustrated in Fig. \ref{eng_spectrum}(a).

Note that applying the Jordan-Wigner transformation to the Hamiltonian
(\ref{heisenberg_hamiltonian}) gives an additional term to Eq. (\ref{diag_h}),
$-1/2(e^{i\theta}c_1^\dagger c_N + e^{-i\theta}c_N^\dagger c_1)(\exp(i\pi\sum_{l=1}^{N}c_l^\dagger
c_l)+1)$, which is a result of the periodic boundary condition. We have already omitted this term
in Eq. (\ref{diag_h}) since it equals zero as long as we consider the one-magnon state. This is
because $\exp(i\pi\sum_{l=1}^{N}c_l^\dagger c_l)+1=0$ for any $N$. Also, the one-magnon condition
makes the Heisenberg model equivalent to the XY model, as the interaction between (virtual)
fermions after the Jordan-Wigner transformation is absent under this condition in both models.

\subsection{Entanglement with an isolated spin}\label{isolated_spin}
\begin{figure}
\begin{center}
\includegraphics[scale=.475]{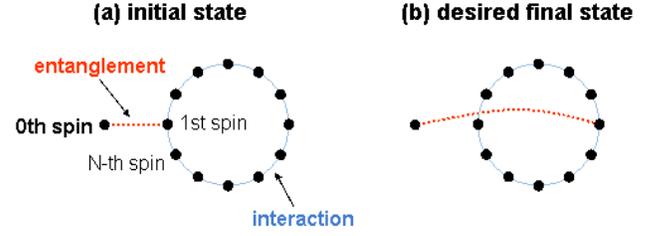}
\caption{Schematic picture of entanglement transfer in a spin chain. (a) The configuration of the
Heisenberg ring and the entanglement it has at $t=0$, with a spatially separated system that is
here represented as the 0th spin. (b) The ideal goal of the entanglement transfer operation: we
wish to transfer as much entanglement with the 0th spin as possible to a specific (target) spin in
the chain.} \label{ent_transfer01}
\end{center}
\end{figure}

Let us analyze the entanglement propagation along the spin chain. Suppose that at $t=0$ a
physically isolated (the 0th) spin and the first spin are maximally entangled as
$(\ket{01}+\ket{10})/\sqrt{2}$ and the rest of the spins in the chain are all in $\ket{0}$. Thus,
the initial state of the whole system can be expressed as
\begin{equation}\label{initial_state}
\ket{\Psi(0)} = \frac{1}{\sqrt{2}}\left[ \ket{0}_0 \frac{1}{\sqrt{N}}\sum_k e^{ik}\ket{k} +
\ket{1}_0 \ket{0}^{\otimes N}\right].
\end{equation}
Hence, we find the state at time $t$, taking $\hbar=1$, as
\begin{eqnarray}\label{psi_at_t}
\ket{\Psi(t)} &=& \frac{1}{\sqrt{2}}\left[ \frac{1}{N}\sum_{k,j}\exp\left(ik(j-1)-i E_k
t\right) \ket{0}_0 S_j^+ \ket{0}^{\otimes N} \right. \nonumber \\
& & \left. + e^{it}\ket{1}_0 \ket{0}^{\otimes N}\right].
\end{eqnarray}
Figure \ref{ent_transfer01} depicts the process we consider: Fig. \ref{ent_transfer01}(a) shows
the initial correlation in $\ket{\Psi(0)}$, and Fig. \ref{ent_transfer01}(b) is the desired goal
of our entanglement transfer operation.

Now we can evaluate the entanglement between the 0th and $l$-th spins. Equation (\ref{psi_at_t})
can be written in the form of Eq. (\ref{one_magnon_state}) with amplitudes
\begin{equation}\label{amp_ac0lth}
\alpha_j(t) = \frac{1}{\sqrt{2}N}\sum_k \exp \left[ik(j-1)- i E_k t\right].
\end{equation}
Because the 0th spin is not interacting with other spins, we can take $|\alpha_0(t)|=1/\sqrt{2}$
for all $t$. From Eq. (\ref{conc_general}) we obtain $C_{0,l}^{0,1}(t) = \sqrt{2} |\alpha_l(t)|$,
where the superscripts on $C$ denote the initially entangled pair. In the limit of large $N$, this
takes a simple analytical form
\begin{equation}\label{conc_bessel_ac}
C_{0,l}^{0,1}(t) = \left| e^{-i(l-1)(\theta-\frac{\pi}{2})} J_{l-1}(t)\right| = \left| J_{l-1}(t)
\right|,
\end{equation}
with the Bessel function of the first kind $J_\nu (x)$. The effect of the AC phase $\theta$
disappears in this limit, since the energy spectrum becomes continuous and displacing all modes by
$\theta$ does not change the overall dispersion relation. This can be clearly seen in the plot in
Fig. \ref{eng_spectrum}(b). In other words, $\theta$ appears only as a common phase factor for all
modes, $e^{-i(l-1)(\theta-\frac{\pi}{2})}$, thus there is no $\theta$-dependence in $|\alpha_j|$.

\begin{figure}
\begin{center}
\includegraphics[scale=.48]{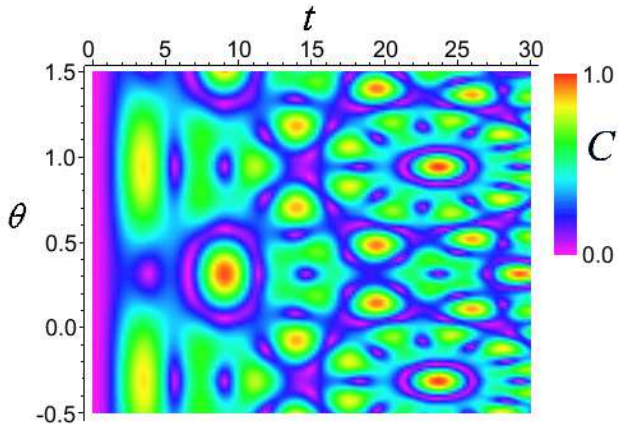}
\caption{An example of plots of the concurrence $C$ transferred in a five-spin chain (ring) in the
presence of a phase shift. The initial entanglement (of the form $(\ket{01}+\ket{10})/\sqrt{2}$)
is in the first spin of the ring and an isolated (0th) spin. The phase shift $\theta$ is an extra
phase a magnon acquires when travelling to a neighboring site. The concurrence $C$ is evaluated
for the pair of the 0th and third spins. Units for $t$ and $\theta$ are $\hbar$ seconds and
radian, respectively, where $\hbar$ is the Planck constant.}\label{N5_3_ac}
\end{center}
\end{figure}

An example of the plots of concurrence as a function of $t$ and $\theta$ is shown in Fig.
\ref{N5_3_ac}, which is the plot of $C_{0,3}^{0,1}(t)$ for $N=5$. The phase $\theta$ is the same
as that in Eq. (\ref{heisenberg_hamiltonian}), that is, the phase a magnon acquires when hopping
from the $j$-th to the $(j+1)$-th site. Some improvement in the amount of transferred entanglement
due to the nonzero phase shift is evident: the maximum concurrence when $\theta=0$ is
$C_{\mr{max}}^{\theta=0}=0.647$ (at $t=59.05$), and when $\theta\neq 0$, $C_{\mr{max}}$ can reach
as high as 0.996 (at $t=23.71$ and $\tan\theta=1.376$) in the region we have calculated, i.e., $t
\in [0,200]$ and $\theta\in [-\pi, \pi]$.

\begin{figure*}
\begin{center}
\includegraphics[scale=.55]{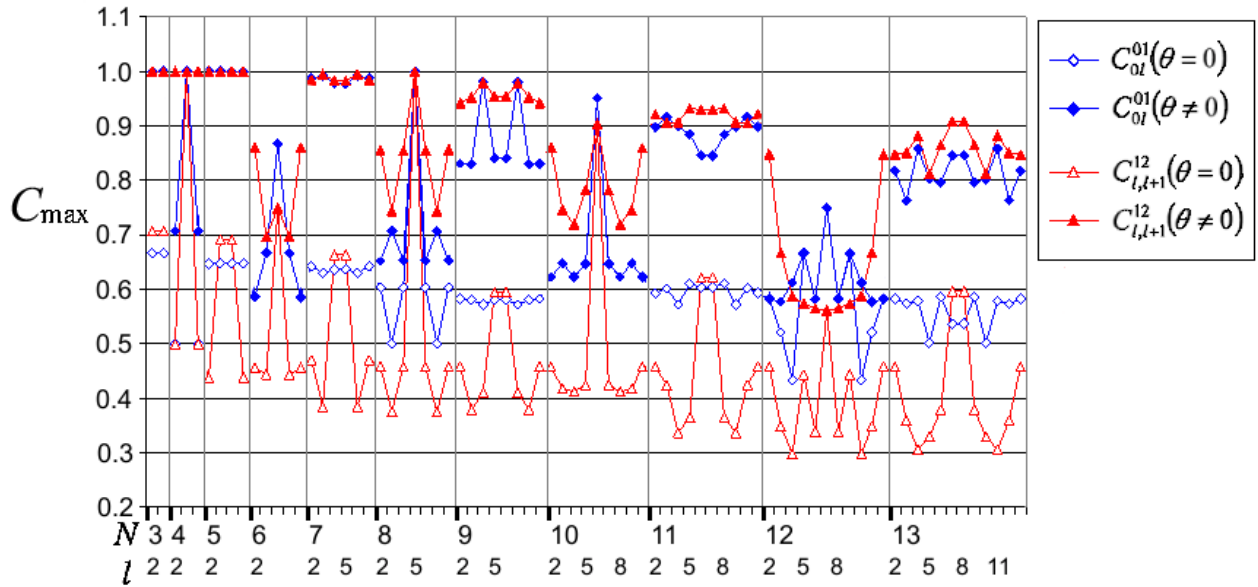}
\caption{Comparison of the maximum concurrence attainable $C_\mr{max}$ between the cases
with/without phase shift. The blue plots (diamonds) show the concurrence between the 0th and
$l$-th spins when the 0th and first spins have the entanglement $(\ket{01}+\ket{10})/\sqrt{2}$ at
$t=0$. The filled and open diamonds correspond to nonzero and zero phase shift, respectively. The
red plots (triangles) are for the concurrence between the $l$-th and $(l+1)$-th spins when the
first and the second spins are initially entangled in the same form. The horizontal axis
represents the total number $N$ of sites in the chain and the location $l$ ($2\le l \le N$) at
which the concurrence is evaluated.}\label{Cmax}
\end{center}
\end{figure*}

Figure \ref{Cmax} shows the comparison between the maximum values of concurrence $C_\mr{max}$ with
and without phase shift for various $(N,l)$ from $(3,2)$ to $(13, 13)$, where $N$ is the total
number of sites and $l$ is the site where the concurrence is evaluated. Plotted are the highest
values found numerically in the range of $0\le t \le 200$. The horizontal axis represents $(N,l)$.
The blue lines with diamond markers are for the geometry with an isolated spin, and the red lines
are for the entanglement transfer when the initial entanglement is held by a pair in the chain.
The latter case will be discussed in the following subsection. For both cases, the open markers
show the maximum concurrence $C_{l_1,l_2}^{m_1,m_2}$ when $\theta=0$, while the filled ones mark
$\mr{max}\{C_{l_1,l_2}^{m_1,m_2}\}$ in the range of $-\pi \le \theta \le \pi$.

The enhancement of entanglement by the phase shift is clearly seen in Fig. \ref{Cmax}. The
transferred entanglement is significantly increased by nonzero phase shifts for many values of
$(N,l)$. Yet, the degree of enhancement varies because quite an effective constructive
interference can occur even when $\theta=0$ for some $(N,l)$.

\subsection{Entanglement in a pair of spins in the chain}
Let us now look at an alternative scenario of entanglement transfer. Instead of being entangled
with an isolated (0th) spin, both spins of an entangled pair can be in the same chain. Considering
the effect of the phase shift mentioned above, we can naturally expect some improvement in the
efficiency of transfer in this scenario as well. We already have the expression Eq.
(\ref{conc_general}) for the concurrence at time $t$ between the $l$-th and $(l+1)$-th spins
$C_{l,l+1}^{m_1,m_2}(t) = 2|\alpha_l^{m_1,m_2} (t)|\cdot |\alpha_{l+1}^{m_1,m_2}(t)|$, where $m_1$
and $m_2$ denote the initial sites that are entangled. Factors $\alpha_j^{m_1,m_2}(t)$ are given
explicitly by
\begin{eqnarray}
\alpha_j^{m_1,m_2}(t) &=& \frac{1}{\sqrt{2}N}\sum_{k} \left(\exp\left[ ik(j-m_1)-i E_k t\right]
\right.
\nonumber \\
& & \left. +\exp\left[ ik(j-m_2)-i E_k t\right]\right).
\end{eqnarray}

The comparison of the maximum concurrence between the cases with and without the phase shift is
shown in Fig. \ref{Cmax} with red triangular markers. For the plot in Fig. \ref{Cmax}, the initial
entanglement of the form of $(\ket{01}+\ket{10})/\sqrt{2}$ is assumed to be in the first and
second spins and the rest are in $\ket{0}$.

As $N$ tends to infinity, $C_{l, l+1}^{m_1,m_2}(t)$ approaches the form
\begin{eqnarray}\label{conc_arb_pairs}
C_{l_1,l_2}^{m_1,m_2}(t) &=& \left| J_{l_1-m_1}(t) + e^{-i(m_1-m_2)(\theta-\pi/2)}
J_{l_1-m_2}(t) \right| \nonumber \\
& & {\hspace{-15mm}} \times  \left| J_{l_2-m_1}(t) + e^{-i(m_1-m_2)(\theta-\pi/2)} J_{l_2-m_2}(t)
\right|.
\end{eqnarray}
Interestingly, unlike Eq. (\ref{conc_bessel_ac}), there is still a dependence on $\theta$,
regardless of $l$'s and $m$'s.  Nonzero $\theta$ can indeed always increase the maximum attainable
pairwise entanglement in a long chain, no matter which pair is initially entangled, and no matter
which pair we evaluate the concurrence for.

We can see in Fig. \ref{Cmax} that the degree of enhancement is larger when the two initially
entangled spins are in the chain, compared with the case of entanglement with an isolated one.
This difference can be understood intuitively when $N$ is large: the physical reasoning for small
$N$'s is essentially the same. If we look at a propagator from a single site, the phase shift
$\theta$ gives an almost common displacement to the phase of all modes at any single site as we
have seen in Eq. (\ref{conc_bessel_ac}). When the entangled pair, $m_1$ and $m_2$, is embedded in
the chain initially, there are two independent propagators stemming from the two sites. Since
these propagators have different phase displacements at the same site, say the $l$-th, the
interference due to non-zero $\theta$ still occurs. As a result, there remains a dependence of the
concurrence on the phase shift.

\section{Spin shift induced by the Dzyaloshinskii-Moriya interaction}\label{dm-type}
Phase shifts can be generated by a different type of interaction, that is, the antisymmetric
exchange interaction, a.k.a. the Dzyaloshinskii-Moriya (DM) interaction
\cite{dzyaloshinskii58,moriya60}, in solids. The DM interaction could be quite significant in some
solid-state-based qubit systems, such as quantum dots \cite{kavokin01}.

The Hamiltonian for the DM interaction between two spins, say 1 and 2, can be written as
\begin{equation}
H_{DM}=\vec{d}\cdot (\vec{S}_1 \times \vec{S}_2),
\end{equation}
where $\vec{d}$ is the coupling vector that reflects the anisotropy of the system. Assuming that
only the $z$-component of $\vec{d}$ has a nonzero value, i.e., $d_z\ne 0, d_x=d_y=0$, and that all
components are constant along the chain, we can write the total interaction Hamiltonian and its
spectrum as (omitting the terms that give only a constant bias)
\begin{eqnarray}\label{dm_hamiltonian}
H &=& -\half \sum_j \left[S_j^+ S_{j+1}^- + S_j^- S_{j+1}^+ +i d_z
(S_j^+ S_{j+1}^- - S_j^- S_{j+1}^+) \right] \nonumber \\
&=& - \,\frac{1}{2\cos\phi} \sum_j (e^{i\phi} S_j^+ S_{j+1}^- + e^{-i\phi} S_j^-
S_{j+1}^+) \nonumber \\
&=& -\,\frac{1}{\cos\phi} \sum_k \cos(k+\phi)\eta_k^\dagger \eta_k,
\end{eqnarray}
where $\phi=\tan^{-1}d_z$. Thus, the difference from the energy spectrum under the AC effect, Eq.
(\ref{energy_spectrum}), is only the factor $1/\cos\phi$ for the energy eigenvalues. As this
factor is independent of $k$, it only rescales the energy spectrum linearly (for a given $\phi$),
hence increases the speed of propagation of all modes by $1/\cos\phi$. Consequently, the maximum
concurrence attainable stays the same as that in the AC effect case, though the time at which the
maximum is achieved should be rescaled as well. All quantitative results in the previous section
are valid for this system with $H_{DM}$ if $t$ is replaced with $t/\cos\phi$.

\section{Summary and Outlook}
We have investigated the effect of externally induced phase shifts on the amount of entanglement
that can be transported in a spin chain. As we have qualitatively anticipated in the Introduction,
these phase shifts can significantly enhance the efficiency of entanglement transfer. Although we
have only studied two shift generating mechanisms, we believe that phase shifts are useful in
quantum information processing, particularly for short-distance transfers, regardless of the
mechanism. Also, we have found that there is an interesting clear difference in the response to
nonzero shift when the chain is sufficiently long.

In the AC-effect-related experiment, there could be a difficulty in providing an electric field
that is intense enough. A rough calculation gives an estimate of the necessary strength of the
electric field of at least $10^7$ V/m to have a meaningful phase shift, if the system size is of
the order of a $\mu$m. Nevertheless, such a strong electric field can be realized by two
dimensional electron gases formed in heterostructured SiGe, GaAs, or other types of III-V
materials. Furthermore, the so-called band-enhancement of spin-orbit coupling in crystals
\cite{andrada97} could be useful to have a substantial AC effect.

Despite a number of technical difficulties, some qubit arrays, in which the effect of the phase
shift can be observed, could be fabricated with present-day technology. For example, consider an
array formed with charge qubits, one type of superconducting qubits \cite{you05}. Quantum
information in a charge qubit is represented by the number of excess Cooper pairs in the
superconducting Cooper-pair box. When neighboring qubits are coupled via a Josephson junction, the
effective interaction can be described by the XY model \cite{siewert00}. As the information
carrier in this case is a pair of electrons, a phase shift can be induced to the wave function of
the pair by the Aharonov-Bohm effect. A magnetic flux $\Phi_B$ threading through a ring formed by
charge qubits with the Josephson-junction-mediated coupling would generate a phase shift $(e/\hbar
c)\Phi_B$ for the wave function of the Cooper pair. Then we could expect the same effect discussed
in this paper. Nonetheless, observing this effect is by no means straightforward: all technical
problems from nano-structure fabrication to measurement method should be addressed. We shall leave
these challenging experimental problems for future investigation.

\begin{acknowledgments}
KM acknowledges Charlie Tahan for stimulating discussions, and Sahel Ashhab for a careful reading
of the manuscript and helpful suggestions. This work was supported in part by the National
Security Agency (NSA), the Laboratory for Physical Sciences (LPS) and the Army Research Office
(ARO); and also by the National Science Foundation (NSF) grant No. EIA-0130383.
\end{acknowledgments}

\section*{Appendix: Note on open ended chains}\label{openendedchains}
As a geometry for information transfer, open ended linear chains may look more natural. If the
dispersion relation can be affected by the phase shift in the case of open ended chain, then the
amount of entanglement transferred can be expected to change as well. However, this is not the
case. Let us briefly look at this.

Suppose that a chain of $N$ spins is placed in a uniform electric field as in Section
\ref{ac-type}. The Hamiltonian is the same as Eq. (\ref{heisenberg_hamiltonian}), but instead of
the periodic boundary condition we have the open boundary condition (OBC), $\alpha_0 =
\alpha_{N+1} =0$. The one-magnon eigenstates are given by
\begin{equation}\label{eigen_openendedchain}
\ket{k}^{\mr{OBC}}=\sqrt{\frac{2}{N+1}}\sum_{j=1}^{N} e^{-ij\theta}\sin (kj) S_j^+
\ket{0}^{\otimes N},
\end{equation}
where $k=\pi n/(N+1)$ with $-N/2< n \le N/2$. Corresponding energy eigenvalues are
\begin{equation}\label{energy_openendedchain}
E_k^{\mr{OBC}}=-\cos k.
\end{equation}
Clearly, the phase shift $\theta$ has no effect on the energy spectrum and thus the propagation
speed of each mode. Therefore $\theta$ causes no change in the concurrence between any two sites.

This result can also be paraphrased in the following way. The phase shifts at all sites can be
cancelled by a product of local gauge transformations, $\Pi_j \exp [ij\theta (S_j^z+1/2) ]$, in
the case of open ended chains. The same cancellation cannot be made for ring-shaped chains because
of the accumulated phase along the chain.

\end{document}